\documentclass[11pt]{article}

\usepackage{euscript}
\usepackage{amssymb}
\usepackage{amsfonts}
\usepackage{amsbsy}
\usepackage{epsfig}
\usepackage{amsthm}
\usepackage{amscd}
\usepackage{amstext}
\usepackage{verbatim}
\usepackage{amsmath}
\usepackage{cancel}
\usepackage{capt-of}
\usepackage{empheq}

\usepackage[pdftex]{hyperref}

\def\ben{\begin{equation}}
\def\een{\end{equation}}

\let\a=\alpha

\let\s=\sigma

\let\w=\omega

\let\pa=\partial
\def\be{\begin{equation}}
\def\ee{\end{equation}}
\def\beq{\begin{equation}}
\def\eeq{\end{equation}}
\def\ba{\begin{array}}
\def\ea{\end{array}}

\def\dalemb#1#2{{\vbox{\hrule height .#2pt
       \hbox{\vrule width.#2pt height#1pt \kern#1pt
               \vrule width.#2pt}
       \hrule height.#2pt}}}

\newcommand{\bea}{\begin{eqnarray}}
\newcommand{\eea}{\end{eqnarray}}

\DeclareRobustCommand{\rchi}{{\mathpalette\irchi\relax}}
\newcommand{\irchi}[2]{\raisebox{\depth}{$#1\chi$}}

\begin{document}

\begin{center}

{ \Large {\bf
Theory of universal incoherent metallic transport
}}

\vspace{1cm}

Sean A. Hartnoll

\vspace{1cm}

{\small
{\it Department of Physics, Stanford University, \\
Stanford, CA 94305-4060, USA }}

\vspace{1.6cm}

\end{center}

\begin{abstract}

In an incoherent metal, transport is controlled by the collective diffusion of energy and charge rather than by quasiparticle or momentum relaxation. We
explore the possibility of a universal bound $D \gtrsim \hbar v_F^2/(k_B T)$ on the underlying diffusion constants in an incoherent metal. Such a bound is loosely motivated by results from holographic duality, the uncertainty principle and from measurements of diffusion in strongly interacting non-metallic systems. Metals close to saturating this bound are shown to have a linear in temperature resistivity with an underlying dissipative timescale matching that recently deduced from experimental data on a wide range of metals.  This bound may be responsible for the ubiquitous appearance of high temperature regimes in metals with $T$-linear resistivity, motivating direct probes of diffusive processes and measurements of charge susceptibilities.

\end{abstract}

\pagebreak
\setcounter{page}{1}

\section*{Introduction}

Many strongly correlated materials exhibit a region in their phase diagram with a linear in temperature resistivity.  Especially ubiquitous are regimes with linear resistivity that extend towards high temperatures. Well known examples include cuprates, pnictides, ruthenates, heavy fermions, vanadium dioxide, fullerenes and organics. It is tempting to search for a universal dynamics that might explain the robustness of this behavior. The obvious challenge is that microscopic details and in particular scattering mechanisms seem to vary significantly between the different materials. However, the notion that a unified framework might capture many of these regimes is supported by the recent experimental observation that a universal dissipative timescale $\tau \sim \hbar/(k_B T)$ underpins the linear in temperature resistivity across a wide range of materials \cite{andy}.

A simple way to make sense of universal behavior is through a bound on some appropriate physical quantity. In strongly correlated systems there are no small parameters and hence they will tend to push up against and saturate any bound. The appeal of this approach is that a bound does not make reference to any particular system and so can aspire to be universal. The difficulty is in identifying the correct kinematic framework in which to formulate the bound. In particular, in metallic systems, the need to relax momentum by processes extrinsic to the electron dynamics (i.e. lattice or disorder scattering) in order to obtain finite conductivities has complicated the search for an intrinsic bound. While the notion of a `planckian' bound on dissipation has been put forward \cite{subir, jan,andy}, a precise proposal for the physical quantity that is bounded is lacking. In this letter we propose that the correct framework for understanding high temperature linear resistivity is that of incoherent metallic transport and that the quantities that are universally bounded are the charge and energy diffusion constants.

In weakly interacting metallic systems with long lived quasiparticles, bounds on transport coefficients can be obtained from the uncertainty relation applied to the mean free path of the quasiparticles. An example of this -- which we review briefly below -- is the Mott-Ioffe-Regel (MIR) resistivity bound, see e.g. \cite{hussey, gun}. However, this bound is violated in strongly correlated `bad metals' \cite{Emery:1995zz}. In bad metals, linear in temperature resistivities cross the MIR bound with impunity. Furthermore, several of the materials exhibiting universal dissipation in \cite{andy} have a resistivity well {\it below} the MIR bound.

For strongly interacting systems one is faced with a paucity of theoretically controlled models. A rare class of strongly interacting theories where dissipative processes can be accurately computed are `holographic' models. Some landmark papers in this endeavor are \cite{Policastro:2001yc, Herzog:2007ij, Hartnoll:2007ih}. Among the most celebrated upshots of holographic studies of dissipation has precisely been evidence for a universal bound on diffusion rates that is approximately saturated by strongly interacting systems. Such a bound was originally conjectured by Kovtun-Son-Starinets (KSS) for the rate of momentum diffusion in a neutral plasma \cite{Kovtun:2004de}, usually described as the ratio of shear viscosity to the entropy density. See \cite{Son:2007vk, Cremonini:2011iq} for overviews of subsequent refinements and developments of the conjecture. A bound on the rate of momentum diffusion is supported by experiments on the strongly correlated quark-gluon plasma and on the Fermi gas at unitarity, see e.g. \cite{Adams:2012th}.

\begin{figure}[h]
\begin{center}
\includegraphics[height = 70mm]{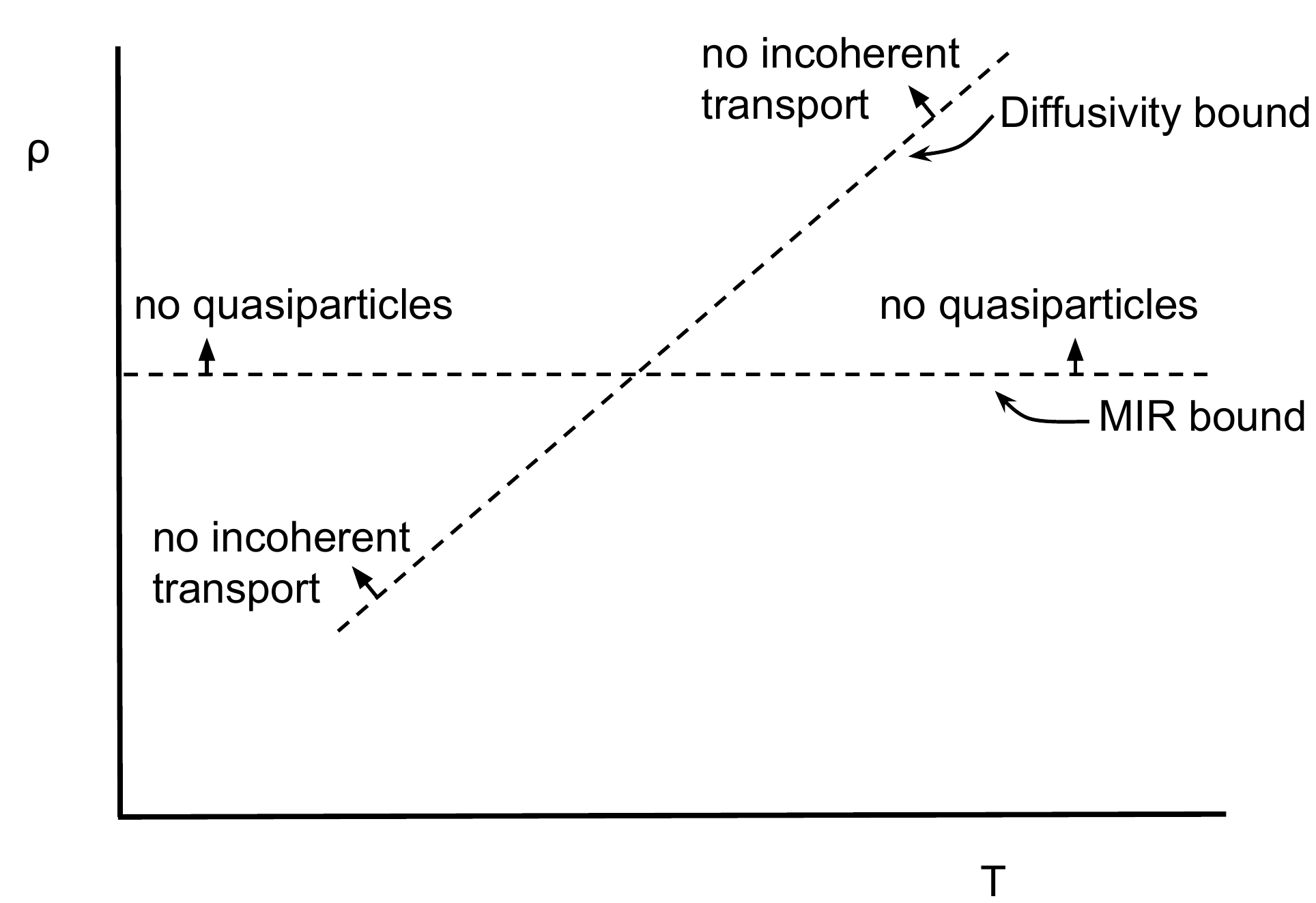}\caption{{\bf Quasiparticle bounds versus incoherent bounds:} Schematic illustration of the role of the Mott-Ioffe-Regel bound and the proposed diffusivity bound in the resistivity versus temperature plane. \label{fig:Dbound}}
\end{center}
\end{figure}

In this letter we propose a universal bound on the diffusivity of charge and energy in an incoherent metal, by analogy to the KSS bound. Momentum and hence viscosity will play no role, by assumption, due to the incoherent nature of the metal. Einstein relations translate these bounds into statements about electric and thermal conductivity. We find that incoherent metals approximately saturating the bound have a linear in temperature electrical resistivity controlled by precisely the timescale $\tau \sim \hbar/(k_B T)$ uncovered in the data \cite{andy}.
Figure \ref{fig:Dbound} below sketches the relation between the bound we consider on incoherent transport and the MIR bound on quasiparticle transport. In strongly interacting systems, the destruction of quasiparticles may set in well below the MIR bound. The incoherence bound is both weaker and stronger than the MIR bound.
In particular, incoherent metals can cross the MIR bound while saturating the diffusivity bound.

\section*{From quasiparticle to incoherent transport}
\label{sec:incoh}

Transport at low and zero frequency is controlled by the longest lived excitations carrying, for example, charge or heat. In a conventional metal, the longest lived excitations are {\it quasiparticles}. The existence of quasiparticles is equivalent to the existence of a large number of almost-conserved (i.e. long-lived) quantities, the quasiparticle densities $\delta n_k$. Each $\delta n_k$ is the number of quasiparticles with momentum $k$ excited above the equilibrium Fermi-Dirac distribution. To describe transport in systems with quasiparticles one needs an equation that describes the interactions and decay of all of these quasiparticles densities. This is why transport in conventional metals is appropriately described by Boltzmann equations \cite{ziman}.

In a system without quasiparticles, one must identify the longest lived excitations. There are two possibilities. We will refer to these as {\it coherent} and {\it incoherent} metals. From this point onwards we will mostly assume that the system of interest does not have long-lived quasiparticles.

In a coherent metal there exists an almost conserved operator (i.e. a long-lived quantity) that overlaps with the current operator. That is, there exists an operator $P$ and total current $J$ such that the susceptibility
\be\label{eq:overlap}
\chi_{JP} \equiv i \int_0^\infty dt \langle [J(t),P(0)] \rangle  \neq 0 \,.
\ee
By long-lived we mean that at late times $\langle P(t) \rangle \sim e^{- \Gamma t}$, with $\Gamma $ a relaxation rate that is parametrically slower than the energy scale at which generic (i.e. non-conserved) quantities decay. For instance $\Gamma \ll k_B T$. It can be shown that the conductivity at small frequencies is then given by a Drude form \cite{forster, Hartnoll:2012rj}
\be\label{eq:drude}
\sigma(\w) = \frac{\chi_{JP}^2}{\chi_{PP}} \frac{1}{- i \omega + \Gamma} \,.
\ee
The simplest way this situation can arise is if momentum-relaxing interactions (umklapp and disorder scattering) remain weak even while momentum-conserving interactions become strong and destroy the quasiparticles. In this case $P$ is the total momentum. It cannot be overemphasized that $\Gamma$ in (\ref{eq:drude}) is the momentum relaxation rate and not a quasiparticle relaxation rate -- there are no quasiparticles by assumption here. Coherent non-quasiparticle transport has recently been studied in \cite{Mahajan:2013cja}, where in particular it was found that such systems have unconventional thermal conduction, with a dramatically low Lorenz ratio $L$. Previous works had emphasized the need to focus on momentum relaxation in strongly interacting metals \cite{rosch, Hartnoll:2007ih, Hartnoll:2008hs, Hartnoll:2012rj}. Holographic models played a useful role in these studies because they are uniquely tractable explicit systems without quasiparticles.

This paper is about incoherent metals. In an incoherent non-quasiparticle metal there will be no long-lived quantities overlapping with the current operators. This implies that there can be no `narrow' Drude peak with a width $\Gamma \ll k_B T$. This still allows for structure in the optical conductivity at low frequencies. For instance, there can be a peak of width $\Gamma \gtrsim k_B T$.
The temperature may be well below the Fermi energy, and the Fermi energy itself may be lower than the interband energy scale \cite{elec}. Such a peak may therefore appear as relatively sharp in the optical conductivity. The important point is that it should not be possible to isolate a relaxation rate from the optical conductivity that is parametrically smaller than the rate at which generic quantities decay due to interactions. This characterizes a regime where both momentum-relaxing and normal interactions are strong, so that there are no quasiparticles and in addition momentum is quickly degraded. Strong microscopic momentum relaxation can appear in low energy effective descriptions as an emergent particle hole symmetry, with $\chi_{JP}$ = 0 in the effective description, as in \cite{fisher, senthil1,senthil2, Metlitski:2010vm}. 

In all metals the conductivity can be expressed in terms of a diffusion constant, via an Einstein relation. In the coherent case, the diffusion constant can be related to the momentum relaxation rate. In these coherent cases it is more convenient to use the Drude-like formula (\ref{eq:drude}) for the conductivity directly. For incoherent metals, however, the Einstein relation is all that one has, and so diffusive processes become central. In the methods section below we derive the generalized Einstein relations
\bea
D_+ D_- & = & \frac{\sigma}{\chi} \frac{\kappa}{c_\rho} \,, \label{eq:e1} \\
D_+ + D_- & = & \frac{\sigma}{\chi} + \frac{\kappa}{c_\rho}
+ \frac{T(\zeta \sigma - \chi \alpha)^2}{c_\rho \chi^2 \sigma} \,. \label{eq:e2}
\eea
These relate the electric, thermal and thermoelectric conductivities -- $\sigma, \kappa, \alpha$ --  to two diffusion constants $D_\pm$, describing the coupled diffusion of charge and heat. The specific heat $c_\rho$, compressibility $\chi$ and thermoelectric susceptibility $\zeta$ are defined in the methods section.

Inhomogeneities in the charge density are effectively instantaneously screened by long range Coulomb interactions. Therefore charge density does not in fact diffuse at the longest wavelengths but rather decays exponentially in time. Nonetheless, as we recall in the methods section below, the Einstein relations between the conductivities and the `bare' (prior to screening) diffusivities still hold. The charge diffusion constant, however, will not be directly measurable through a physical diffusion processes. The energy density, in contrast, is not screened and hence the thermal diffusivity can be directly measured in principle.

\section*{Bounds on transport coefficients}
\label{sec:bounds}

Within the framework of quasiparticle transport, it is possible to bound transport coefficients. We can first describe the 
Mott-Ioffe-Regel (MIR) limit. From the Drude formula for a $d$-dimensional metal
\be\label{eq:MIR}
\sigma = \frac{n \tau e^2}{m} \sim (k_F \ell) \,  k_F^{d-2} \frac{e^2}{\hbar} \, \gtrsim \, k_F^{d-2} \frac{e^2}{\hbar} \,.
\ee
Here we imposed $k_F \ell \gtrsim 1$; the mean free path of the quasiparticles should be longer than their Compton wavelength.  Equation (\ref{eq:MIR}) is a version of the MIR bound, and implies an upper bound on the resistivity of a metal. Violation of the MIR limit is widely observed in strongly correlated systems at high temperatures, suggesting that transport in these `bad metal' regimes is not controlled by quasiparticle physics \cite{hussey, gun, Emery:1995zz}. In addition, strongly correlated systems can have resistivities well below the MIR bound (e.g. \cite{andy}). This suggests that the MIR bound is both too strong and too weak.

The MIR limit above was stated in terms of the mean free path of the quasiparticles, but from our perspective the lifetime is the more fundamental property. So let us instead apply the energy-time uncertainty principle to the quasiparticle lifetime $\tau$. The quasiparticles have energy $k_B T$ and therefore we obtain
\be\label{eq:strong}
\sigma = \frac{n \tau e^2}{m} \sim \frac{\tau E_F}{\hbar} k_F^{d-2} \frac{e^2}{\hbar} \, \gtrsim \, k_F^{d-2} \frac{E_F}{k_B T} \frac{e^2}{\hbar} \,.
\ee
This result requires sufficiently inelastic interactions for energy to equilibrate. At low temperatures, elastic impurity scattering leads to a residual resistivity violating (\ref{eq:strong}). Throughout we work at high temperatures away from the residual resistivity and any localization or insulating physics.
For temperatures $k_B T \lesssim E_F$ we see that the standard MIR bound (\ref{eq:MIR}) is weaker than it could be.
The stronger bound (\ref{eq:strong}) now implies that a quasiparticle-based resistivity is bounded above by $T$-linearity. There is at least one case in which quasiparticles can saturate this stronger bound. Scattering of electrons by phonons at temperatures above the Debye scale, with an order one electron-phonon coupling, gives a conductivity saturating (\ref{eq:strong}). Due to Migdal's theorem for electron-phonon interactions, the quasiparticles are not destroyed \cite{kad2}. For more general interactions, such as fermions interacting with gapless critical bosons, quasiparticles will not survive with $\tau \sim \hbar/(k_B T)$.

The same uncertainty principle argument -- again applied to systems with long lived quasiparticles -- has been made to bound the shear viscosity divided by the entropy density in a plasma \cite{Kovtun:2004de}. Here
\be\label{eq:etas}
\frac{\eta}{s} \sim \frac{\epsilon \tau}{k_B n} \, \gtrsim \, \frac{\hbar}{k_B} \,.
\ee
In the above expression $\epsilon/n$ is the energy per quasiparticle, and this is the quantity used in the uncertainty relation. Shear viscosity quantifies the diffusion of momentum (e.g. \cite{Son:2007vk}). In a relativistic system with vanishing chemical potential, in particular, $\eta/s = D \, T/c^2$. Here $D$ is the momentum diffusion constant and $c$ is the speed of light. Therefore we can rewrite (\ref{eq:etas}) as
\be\label{eq:Dbound}
\frac{D}{c^2} \, \gtrsim \, \frac{\hbar}{k_B T} \,.
\ee
It is important to stress that the above bound only follows from (\ref{eq:etas}) in the relativistic case with vanishing chemical potential. We are instead interested in a finite charge density, non-relativistic limit. Nonetheless,
the above expression (\ref{eq:Dbound}) resonates with the phenomenology of strongly interacting `planckian' \cite{jan} quantum critical systems in which temperature is the only scale \cite{subir}.

Bounds such as (\ref{eq:etas}) and (\ref{eq:Dbound}) can be stated without any reference to an underlying quasiparticle description. This fact, combined with the observation that a wide range of (non-quasiparticle) theories with holographic gravity duals had a low but specific value of $\eta/s$, led Kovtun, Son and Starinets (KSS) to conjecture a universal bound for all systems \cite{Kovtun:2004de}
\be\label{eq:KSS}
\frac{\eta}{s} \, \geq \, C \frac{\hbar}{k_B} \,.
\ee
The KSS bound was stated with a specific constant $C$. Subsequent work found counterexamples to the original statement, but the powerful notion that a specific (to be discovered) $C$ does exist such that (\ref{eq:KSS}) is always true has survived \cite{Cremonini:2011iq}. The key conjecture is that in some suitably generic class of systems -- {\it with or without quasiparticles} -- the dissipation timescale is fundamentally bounded. The evidence for this conjecture beyond the quasiparticle regime comes both from holographic theories, see e.g. \cite{Son:2007vk, Cremonini:2011iq}, and from measurements of the viscosity in strongly interacting systems such as the quark-gluon plasma and the Fermi gas at unitarity, see e.g. \cite{Adams:2012th}. On the other hand, one measurement of transverse spin diffusion in a strongly interacting cold atomic Fermi gas has measured a very low diffusivity, challenging proposed bounds \cite{transverse}. To obtain a sharp contradiction, the (currently unknown) renormalized mass and Fermi energy for this system are needed.

It is not straightforward to apply the viscosity bound to transport in metals (see \cite{Davison:2013txa,Lucas:2014zea} for a recent interesting discussion). The problem is the following: the conductivities in a strongly interacting coherent metal are determined by the momentum relaxation rate $\Gamma$, not by momentum diffusion. Momentum diffusion is an intrinsic property of the system, described by the diffusion constant $D$. Momentum relaxation is instead determined by the way in which the system is coupled to external translation symmetry breaking effects such as disorder or a lattice \cite{Hartnoll:2007ih, Hartnoll:2008hs, Hartnoll:2012rj}. Thus momentum relaxation effects are extrinsic to the electronic system and are not generally universal. An exception occurs when the momentum degrading dynamics involves only long wavelength processes that can themselves be described hydrodynamically \cite{ks1, ks2, Balasubramanian:2013yqa}. It also occurs in certain holographic systems with dynamical critical exponent $z=\infty$ \cite{Davison:2013txa}.

In this letter we are arguing that transport in incoherent metals is described by diffusive physics. Not by momentum diffusion (and hence not related to the viscosity) but rather by diffusion of charge and energy. As we have said, diffusion is intrinsic to the system. Therefore, we would like to suggest that incoherent metals are precisely the metallic systems to which universal bounds on the rate of dissipation can be usefully applied. Specifically, it is tempting to adapt the diffusive bound (\ref{eq:Dbound}) to our diffusion constants $D_\pm$ appearing in (\ref{eq:e1}) and (\ref{eq:e2}) as
\be\label{eq:dcharge}
\frac{D_\pm}{v_F^2} \, \gtrsim \, \frac{\hbar}{k_B T} \,.
\ee
We have replaced the speed of light appearing in (\ref{eq:Dbound}) with the characteristic speed in a metal: the Fermi velocity $v_F$. This is a nontrivial sleight of hand. In particular, the KSS bound is only equivalent to (\ref{eq:Dbound}) for relativistic systems with zero charge density. In some regards, however, $v_F$ does play a role analogous to the speed of light -- for instance by mediating a linear dispersion relation for the low energy excitations of the Fermi surface. Furthermore, the diffusion constants in a (quasiparticle based) Fermi liquid satisfy $D_\text{qp} \sim v_F^2 \tau$; the uncertainty principle arguments above then imply 
\be
\frac{D_\text{qp}}{v_F^2} \gtrsim \frac{\hbar}{k_B T} \,,
\ee
in agreement with our more general proposed bound (\ref{eq:dcharge}). This is of course effectively the same quasiparticle bound as we obtained for the conductivity in (\ref{eq:strong}). As we have noted, in a quasiparticle regime there is no distinction between coherent and incoherent transport. The virtue of rephrasing the quasiparticle bound in terms of diffusion as in (\ref{eq:dcharge}) is that it can now be continuously applied to strongly interacting incoherent metals also, sidestepping extrinsic momentum relaxation. With strong interactions the Fermi velocity may not be unambiguously defined. As a working notion, we can imagine using $v_F$ extracted from quantum oscillations at very low temperatures, as in \cite{andy}. A more precise statement of (\ref{eq:dcharge}) is clearly desirable, ideally without reference to any single-particle properties. The strongest argument for (\ref{eq:dcharge}) at present may be the interesting phenomenology it leads to.

The bounds on the diffusion constants in (\ref{eq:dcharge}) translate into bounds on the conductivities through (\ref{eq:e1}) and (\ref{eq:e2}). This translation is complicated by the thermoelectric conductivity and susceptibility -- $\alpha$ and $\zeta$ -- appearing in (\ref{eq:e2}). In many materials of interest these thermoelectric terms are expected or measured to be small by powers of $T/E_F$ compared to the other terms on the right hand side of (\ref{eq:e2}).  For simplicity, in this first pass, we proceed to drop these thermoelectric terms and thereby conclude that
\be\label{eq:universal}
\frac{\sigma}{\chi} \,, \frac{\kappa}{c_\rho} \; \gtrsim \; \frac{v_F^2 \, \hbar}{k_B T} \,.
\ee
The bounds in (\ref{eq:universal}) underpin much of our remaining discussion.

Incoherent metals have been widely studied using dynamical mean field theory (DMFT), see e.g. \cite{dm1,dm2}. This approach may help provide a microscopic picture of the dynamics underlying bounds such as (\ref{eq:dcharge}). In fact, we can quickly reproduce a result from the ultra high temperature expansion of DMFT just from our Einstein relations. In a two dimensional metal at temperatures well above the bandwidth $E_B$ one obtains (from the Fermi-Dirac distribution) the susceptibilities $c_\rho \sim E_B^3/T^2$ and $\chi \sim E_B/T$. The diffusion constants are set by the lowest energy scale, which is now $E_F$. So $D \sim 1/E_F$ is temperature independent and clearly compatible with the bound (\ref{eq:dcharge}) at high temperatures. From the Einstein relations (neglecting thermoelectric conductivities) we then obtain the conductivities $\sigma \sim 1/T$ and $\kappa \sim 1/T^2$. These are the results obtained in DMFT in \cite{dmft}, which are now seen to be general properties of incoherent diffusion in the ultra high temperature limit.

A bound of the form (\ref{eq:universal}) was discussed in \cite{Kovtun:2008kx}, with $v_F \to c$, for the electrical conductivity of a conformal field theory (CFT). In a CFT, particle-hole symmetry implies $\chi_{JP} = 0$ in (\ref{eq:overlap}). Hence the d.c.~conductivity is indeed controlled by universal charge diffusion rather than momentum relaxation \cite{ds, subir}. In a CFT with two space dimensions, for example, the conductivity $\sigma$ is constant and the temperature dependence for (\ref{eq:universal}) is supplied by the charge susceptibility $\chi$. Here we are instead discussing metallic systems in which  strong microscopic momentum relaxation is required to enter a diffusive, incoherent regime. The proposal is that (\ref{eq:universal}) holds for this restricted class of incoherent metals.

\section*{Experimental consequences}

We proceed to describe simple consequences of the proposed universal bounds (\ref{eq:dcharge}) and (\ref{eq:universal}) on transport in incoherent metals. We can expect that strongly correlated systems with no small coupling constants will tend to push up against the bound, up to order one numerical factors. Further assuming that we can drop the thermoelectric terms in (\ref{eq:e2}), we describe the phenomenology of incoherent metals that approximately saturate the bounds (\ref{eq:universal}).

Upon saturating the bound (\ref{eq:universal}), the electrical resistivity is linear in temperature:
\be\label{eq:rho}
\rho \, \sim \, \frac{1}{v_F^2 \chi} \frac{k_B T}{\hbar} \, \sim \,  \frac{\hbar}{k_F^{d-2} E_F} \frac{k_B T}{e^2} \,.
\ee
Here and throughout we will be discussing conductivities in the number of conducting dimensions $d$. Thus, for example, to get the predicted resistivity for a metal with two dimensional conducting layers, the expression in (\ref{eq:rho}) should be multiplied by the interlayer spacing. In the second relation in (\ref{eq:rho}) we have estimated the charge susceptibility $\chi \sim e^2 k_F^d/E_F$, and also $E_F \sim \hbar k_F v_F$. In particular, we are assuming the susceptibility does not have a strong temperature dependence. As with the Fermi velocity, the Fermi momentum $\hbar k_F$ may not be sharply defined without quasiparticles. We can again imagine that these quantities have been extracted from low temperature quantum oscillations. If we (na\"ively and incorrectly in a non-quasiparticle system) match the expression for the resistivity (\ref{eq:rho}) to a Drude formula, we will extract the effective timescale
\be\label{eq:timescale}
\tau_\text{eff.} \, \sim \, \frac{\hbar}{k_B T} \,.
\ee
Precisely this timescale (up to an order one numerical prefactor that cannot be discussed at the accuracy we are working) indeed underpins the linear in temperature resistivity in a large swath of materials including cuprates, ruthenates, heavy fermions, organics and elemental metals \cite{andy}. The framework of incoherent metals developed here is therefore a candidate for understanding the ubiquity of this `planckian' \cite{jan, subir} linear in temperature resistivity. The fundamental requirements are that (i) momentum is relaxed quickly and therefore that (ii) transport is controlled by diffusion of energy and charge that is subject to `universal' bounds.

Some of the materials described in \cite{andy} -- in particular the elemental metals -- are likely not incoherent metals of the sort we are describing. The scattering timescale (\ref{eq:timescale}) in these cases presumably arises through standard `Prange-Kadanoff' \cite{kad2} quasiparticle scattering by classicalized phonons, together with an order one electron-phonon coupling.  Below we will describe clear signatures that distinguish these two sources of planckian dissipation. Firstly, scattering of quasiparticles by classicalized modes entails the Wiedemann-Franz law \cite{ziman, Mahajan:2013cja}. This law will not hold for incoherent metals. Secondly, scattering by quasiparticles cannot cross the MIR bound, while incoherent transport obeying (\ref{eq:universal}) can do so.

Universal incoherent transport offers a perspective on the challenge of `bad metals' \cite{Emery:1995zz}, whose resistivity violates the MIR limit. From (\ref{eq:rho}) we have
\be\label{eq:rratio}
\frac{\rho}{\rho_\text{MIR}} \sim \frac{k_F^d}{E_F^2 \chi} k_B T \sim \frac{k_B T}{E_F} \,.
\ee
Therefore we see that universal incoherent metals cross the MIR bound when $k_B T \sim E_F$. This can happen at experimentally reasonable temperatures if $E_F$ is relatively low (as it tends to be in strongly correlated metals, as measured by e.g. the spectral weight \cite{elec,basov}) and with the help of some order one numerical factors in (\ref{eq:timescale}) or multiple Fermi surfaces. Of course, many of the estimates we are employing will require a more careful treatment in order to be applied to systems with multiple or highly anisotropic Fermi surfaces. We see in (\ref{eq:rratio}) that a small charge susceptibility $\chi$ helps to lower the temperature at which the MIR bound is reached. This observation motivates the experimental characterization of charge susceptibilities in bad metals. The charge susceptibility is also important in relation to possible nearby Mott transitions \cite{MIT, MIT2, MIT3}.
The main observation, however, is that incoherent metals saturating the proposed diffusivity bounds (\ref{eq:universal}) can happily cross the MIR limit at high enough temperatures and become bad metals. See figure \ref{fig:Dbound} above. The strongly correlated materials analyzed in \cite{andy} include cases that cross the MIR bound (a pnictide and a cuprate) and cases that do not (heavy femions, an organic material and Sr$_3$Ru$_2$O$_7$).

A key feature of an incoherent metal is the absence of a Drude peak with a width narrower than $\Gamma \sim \min \left\{k_B T, E_F \right\}$. This is a distinctive feature that differentiates incoherent metals from a strongly coupled coherent metal. Any structure at small frequencies in the optical conductivity must broaden as the temperature is increased, and by the time the MIR bound is reached it must have broadened out to $\Gamma \sim E_F$. Precisely this behavior is well documented in the cuprate and pnictide superconductors. Optical conductivity data at high temperatures in LSCO shows the disappearance of the Drude peak \cite{takenaka,hussey}. Substantial broadening of the Drude peak at increasing temperatures is also seen in the $T$-linear resistivity regimes of, for instance, Bi-2212 \cite{marel,hwang} and YBCO \cite{boris} cuprates, and in pnictides \cite{p1,p2}. Further instances of metals with $T$-linear resistivity above the MIR bound and with beautifully incoherent optical conductivities include the organic metal $\theta$-(BEDT-TTF)$_2$I$_3$ \cite{i1}, LiV$_2$O$_4$ \cite{i2}, Na$_{0.7}$CoO$_2$ \cite{i3} and CaRuO$_3$ \cite{i4}. The optical conductivities of these materials do not even have Drude peaks at high temperatures. Instead the peak in the optical conductivity moves away from $\w = 0$ as it broadens. In holographic studies of strongly correlated media, such broad peaks are typically associated with `quasinormal modes', see e.g. \cite{WitczakKrempa:2012gn}. It seems clear that this behavior cannot be explained by quasiparticle scattering by a classicalized `phonon'-type mode. All of these materials are excellent candidates for systems exhibiting universal incoherent metallic transport.

The onset of incoherent metallic transport at high temperatures is often discussed in terms of spectral weight transfer \cite{hussey, elec}. The spectral weight in the Drude peak is transferred to very high energy scales as the metal becomes increasingly incoherent. This has led to recent suggestions that the $T$-linear resistivities and bad metallic behavior should be understood as due to a strong temperature dependence of the plasma frequency $\omega_p$ (or some suitably defined effective plasma frequency $\omega_p^*$), rather than an anomalous scattering rate $\tau$, e.g. \cite{kot,nickel}. These analyses are not in contradiction with the framework we are developing. In particular, it is possible that approaches such as DMFT may help give a microscopic understanding of universal incoherent transport. From the perspective of incoherent transport, however, the quantities that organize the temperature dependence of the resistivity are the diffusion constant $D$ and charge susceptibility $\chi$, via $\sigma = D \chi$. The quantities $\tau$ and $\omega_p$ are less natural in the context of diffusive conduction. There are indeed different ways in which the charge diffusivity and susceptibility can scale to give a $T$-linear resistivity: we have focussed on $\{D \sim 1/T, \chi \sim 1\}$ in the context of saturating the bound (\ref{eq:dcharge}), but another possibility is $\{D \sim 1, \chi \sim 1/T\}$, which is relevant in the ultra high temperature limit. Another possibility compatible with the bound (\ref{eq:dcharge}) at $k_B T \lesssim E_F$ is the scaling $\{D \sim E_F/T^2, \chi \sim T\}$, although this last case would presumably have long-lived quasiparticles (cf. \cite{kot}) and hence is not properly incoherent. Interestingly, the system could move between these different behaviors with the dc conductivity remaining $T$-linear. To clearly distinguish these regimes, it would be extremely interesting to directly measure the diffusion constant.

A further prediction of (\ref{eq:rho}) is that the coefficient of the linear in temperature resistivity should vary inversely with changes in $E_F$. In underdoped LSCO it has been observed that at high temperatures the coefficient of the $T$-linear resistivity decreases as the doping $x$ is increased towards optimal doping \cite{ando,hussey1}. And, indeed, $E_F$ also appears to increase with $x$ towards optimal doping, as measured from the spectral weight \cite{orenstein,uchida}. These observations are therefore qualitatively in agreement with (\ref{eq:rho}). In discussing the coefficient of the linear in T resistivity, it is important to distinguish the high temperature incoherent regime, of interest to us here, from the linear in T resistivity appearing at low temperatures upon application of large magnetic fields \cite{hussey1}. The strong doping dependence of the coefficient of the low temperature $T$-linear resistivity, in particular on the overdoped side, appears to be closely tied to the onset of superconductivity \cite{taillefer}.

Scattering by classicalized modes such as phonons or magnetic fluctuations, above some effective Debye scale and with an effective coupling of order one, is a competing mechanism to produce the dissipative timescale (\ref{eq:timescale}).
This is a quasiparticle mechanism and so cannot be applied above the MIR limit. This mechanism therefore seems to be excluded for the cuprates, pnictides and other bad metals we mentioned above (at least above the MIR bound). A distinctive feature of this mechanism is that the electrons experience the scattering as elastic because the modes they scatter off all necessarily have lower energies, below the effective Debye scale. This leads \cite{ziman, Mahajan:2013cja} to the Wiedemann-Franz law holding: $L \equiv \kappa/(\sigma T) = \pi^2/3 \, \times \, k_B^2/e^2$. However, it has been observed that the Wiedemann-Franz law does {\it not} hold in the `high' temperature regime with $T$-linear resistivity in heavy fermions \cite{tanatar,con1} and in a cuprate \cite{zhang}. This suggests that these
particular $T$-linear resistivities are not due to scattering off classicalized modes \cite{Mahajan:2013cja}.
In contrast, in an incoherent metal the Wiedemann-Franz law will not hold. 
In particular, the electronic Lorenz ratio is given by
\be\label{eq:L}
L \equiv \frac{\kappa}{T \sigma} \, \approx \, \frac{c_\rho}{\chi \, T} \, \sim \, \frac{k_B^2}{e^2} \,.
\ee
To obtain the final term we used the estimate $c_\rho \sim k_F^d/E_F \times T k_B^2$. The phonon contribution to the thermal conductivity must be subtracted off, for instance by measuring the thermal Hall conductivity \cite{zhang}.
By assuming that the diffusion constants $D_+ \approx D_-$ are comparable, the ratio of conductivities has become a `Wilson ratio' of susceptibilities. 

It is interesting to estimate the temperature dependence of the Lorenz ratio (\ref{eq:L}) compared to the Fermi liquid value of $L_0 \equiv \pi^2/3 \, \times \, k_B^2/e^2$.
For a rough thermodynamic estimate we take a Fermi-Dirac distribution of free electrons at chemical potential $\mu$ and temperature $T$. We also include a cutoff on the energy integral at some bandwidth scale $E_B$. The susceptibilities are easily computed from the free energy; e.g. for a two dimensional metal $f = - g_0 T \int_0^{E_B} d\epsilon \log \left(1 + e^{-(\epsilon - \mu)/T} \right)$. The ratio of susceptibilities is shown in figure \ref{fig:WF} below.
\begin{figure}[h]
\begin{center}
\includegraphics[height = 60mm]{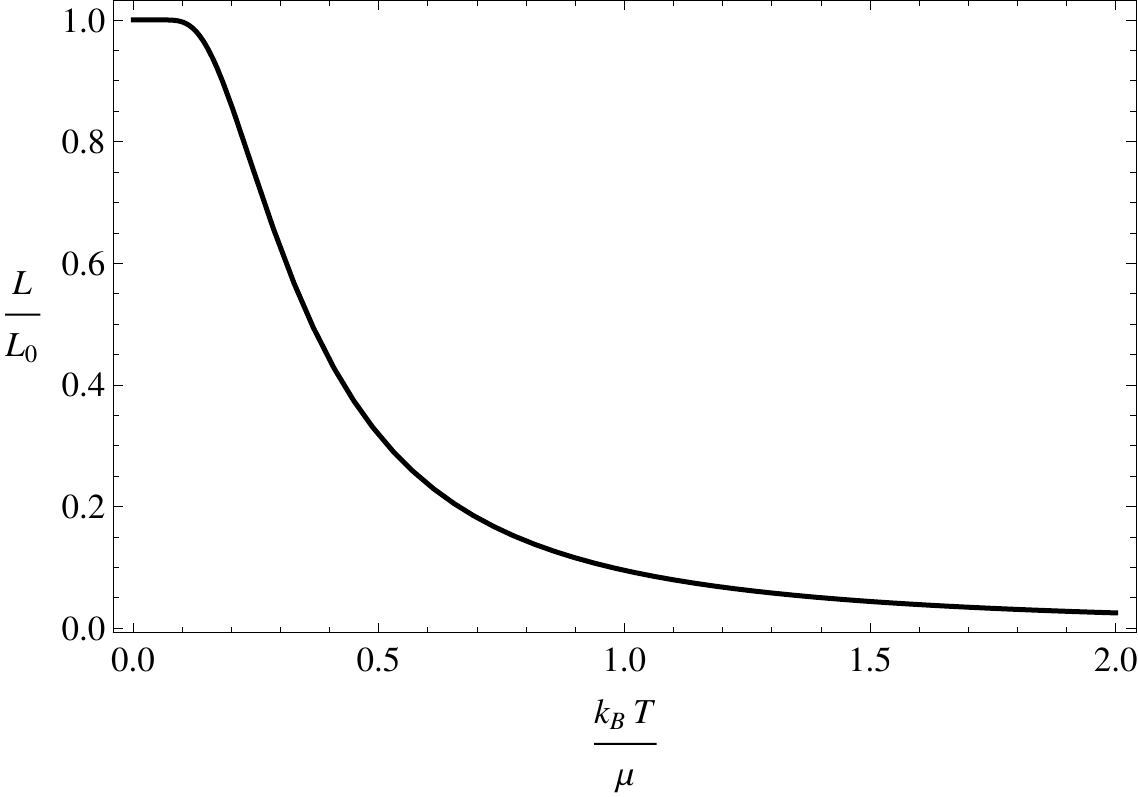}\caption{{\bf Lorenz ratio as a function of temperature.} Estimated using a Fermi-Dirac distribution for two dimensional electrons to compute the thermodynamic susceptibilities in (\ref{eq:L}). Temperature $T$, chemical potential $\mu$ and bandwidth cutoff on the single particle energy taken (for illustrative purposes) to be $E_B = 2 \mu$.  \label{fig:WF}}
\end{center}
\end{figure}
The important result contained in this figure is that the Lorenz ratio in an incoherent metal can drop dramatically from the  Fermi liquid value of $L_0$, even at temperatures that are relatively low compared to the chemical potential scale. At high temperatures above $E_B$ we cross over to the ultra high temperature behavior $L \sim 1/T^2$, as follows from the results below equation (\ref{eq:universal}) above and also obtained from DMFT in \cite{dmft}. This is a distinct mechanism to achieve a very low $L$ than that found in \cite{Mahajan:2013cja} for coherent metals.

Universal incoherent metallic transport captures many observed features of high temperature regimes with $T$-linear resistivity. The most compelling evidence that a given material is an incoherent metal in the sense we have described would, however, be the direct detection of diffusive processes. This requires the measurement of spatially resolved dissipative dynamics as described by e.g. equation (\ref{eq:wk}) in the methods section. It is of interest to directly extract the diffusion constants $D_\pm$, perhaps through time and space resolved measurements of photoresponse. As we have noted, charge diffusion cannot be directly observed due to screening. Thermal diffusion does not suffer from this problem -- indeed,
a photothermal signal has been used to measure the thermal diffusivity of BSCCO in the incoherent $T$-linear resistivity regime \cite{d1,d2}. The data shows a slow decrease in the diffusivity as the temperature is increased. Extracting the electronic contribution as discussed in \cite{d1} leads to an electronic thermal diffusion constant of the order of the proposed bound (\ref{eq:dcharge}). It is reasonable to expect that $D_+ \sim D_-$, but analogous independent determination of the unscreened charge diffusivity, perhaps through studying the plasmon mode or by measuring the compressibility, is desirable. Other measurements of diffusion in metals include spin diffusion \cite{spin} and quasiparticle diffusion \cite{qp} in the superconducting state. Detailed characterization of $D_\pm$, ideally with the temperature dependence predicted by (\ref{eq:dcharge}), could strongly corroborate the picture of incoherent transport that we have explored here.

\section*{Final comments}

We have proposed the bounds (\ref{eq:dcharge}) on the diffusive processes that control transport in an incoherent metal. The bounds were introduced by analogy with those that have been conjectured (with some theoretical and experimental evidence) for other diffusive processes, notably momentum diffusion. However, there was a certain logical leap from existing bounds to (\ref{eq:dcharge}). The precise role of the Fermi velocity in the bound as well as the precise set of systems to which it can be expected to apply need to be clarified. Controlled holographic models of incoherent metals may help with these questions. Incoherent metals have only very recently been found in holographic theories  \cite{Donos:2014uba, Gouteraux:2014hca}. Incoherent metallic transport is also achieved holographically through probe brane systems \cite{Karch:2007pd, Hartnoll:2009ns}, although in these cases energy as well as momentum is also strongly relaxed into a bath.

In going from (\ref{eq:dcharge}) to (\ref{eq:universal}) we dropped terms depending on the thermoelectric conductivity and susceptibility on the grounds that they were suppressed by powers of $k_B T/E_F$. As we have noted, this ratio may not be especially small in bad metallic regimes. Non-Fermi liquids may also have anomalously large thermopower. A more complete treatment of these terms (and, ideally, measurements of $\alpha$ and $\zeta$ in the relevant regimes) is desirable.

The above caveats aside, the main achievement of our discussion has been to identify a regime where universal bounds on strongly correlated transport, intrinsic to the metallic dynamics, can be usefully proposed. We reviewed strong evidence from optical conductivity that many interesting materials are indeed in an incoherent metallic regime over the relevant regions of their phase diagram. Among other phenomenologically appealing features, the proposed universal incoherent metallic transport is largely insensitive to the Mott-Ioffe-Regel bound, as we illustrated in figure \ref{fig:Dbound} above. It is hoped that thinking in terms of plausible universal bounds can help demystify some of the anomalous yet ubiquitous characteristics of these materials.

\section*{Methods}
\label{sec:diff}

\subsection*{Diffusion of heat and charge in a metal}

The total energy and the total electric charge in the system are certainly conserved. Local fluctuations of the energy and charge densities, $\epsilon$ and $\rho$, therefore obey
\be\label{eq:conserved}
\frac{\pa \epsilon}{\pa t} + \nabla \cdot j^E = 0 \,, \qquad \frac{\pa \rho}{\pa t} + \nabla \cdot j = 0 \,. 
\ee
Here $j$ is the electric current and $j^E$ is the energy current. Long wavelength fluctuations of the energy and charge densities will be the longest-lived modes in an incoherent metal. These are scalar modes and therefore cannot directly overlap with the vectorial currents in the sense of (\ref{eq:overlap}) in the infinite wavelength (zero momentum) limit. However, the diffusive dynamics of the conserved densities does in fact control charge and heat transport at late times and therefore determines the conductivities. To describe this effect we need a slight generalization of the Einstein relation between diffusivity and mobility. At long wavelengths, gradients of the temperature and chemical potential produce currents according to
\be\label{eq:fick}
j = - \sigma \nabla \mu - \alpha \nabla T \,, \qquad j^Q = - \alpha T \, \nabla \mu - \overline \kappa \, \nabla T \,.
\ee
Here the heat current
\be\label{eq:jQ}
j^Q = j^E - \mu j \,,
\ee
and the conductivities $\sigma,\alpha,\overline \kappa$ are constant. The two `thermoelectric' conductivities are related as usual by Onsager reciprocity.

Fluctuations of the densities are related to fluctuations of the temperature and chemical potential through thermodynamic susceptibilities. At fixed chemical potential and temperature, the thermodynamic potential density is $f = \epsilon - s T - \mu \rho$. Define
\be
\chi \equiv - \frac{\pa^2 f}{\pa \mu^2} \,, \qquad c_\mu \equiv - T \frac{\pa^2 f}{\pa T^2} \,, \qquad
\zeta \equiv - \frac{\pa^2 f}{\pa T \, \pa \mu} \,.
\ee
Derivatives are taken with $\mu$ or $T$ fixed as appropriate. Then, from elementary thermodynamics, we have
\be\label{eq:change}
\nabla \rho = \chi \nabla \mu + \zeta \nabla T \,, \qquad \nabla \epsilon = \left(T \zeta + \mu \chi \right) \nabla \mu + \left(\mu \zeta + c_\mu \right) \nabla T \,.
\ee
Putting equations (\ref{eq:conserved}), (\ref{eq:fick}), (\ref{eq:jQ}) and (\ref{eq:change}) together it is clear that we will obtain two coupled diffusion equations for $n_A \equiv \{ \epsilon, \rho \}$:
\be\label{eq:dif}
\frac{d}{dt} n_A = D_{AB} \nabla^2 n_B \,.
\ee
These equations are easily decoupled by diagonalizing the matrix $D$ of diffusion constants. The eigenvalues $D_{\pm}$ of this matrix satisfy
\bea
D_+ D_- & = & \frac{\sigma}{\chi} \frac{\kappa}{c_\rho} \,, \label{eq:e1aa} \\
D_+ + D_- & = & \frac{\sigma}{\chi} + \frac{\kappa}{c_\rho}
+ \frac{T(\zeta \sigma - \chi \alpha)^2}{c_\rho \chi^2 \sigma} \,. \label{eq:e2aa}
\eea
Here we introduced the thermal conductivity with open circuit boundary conditions (the usual thermal conductivity)
\be
\kappa = \overline \kappa - \frac{\a^2 T}{\s} \,,
\ee
as well as the specific heat at fixed charge density
\be
c_\rho = c_\mu - \frac{\zeta^2 T}{\chi} \,.
\ee
The above derivation can be streamlined by working with entropy rather than energy diffusion. In one dimension, nonlinear corrections to (\ref{eq:dif}) can be important \cite{vadim}.

The expressions (\ref{eq:e1aa}) and (\ref{eq:e2aa}) are the generalized Einstein relations. In the absence of any coupling between the charge and heat carriers (for instance, in a particle-hole symmetric system) then $\zeta = \alpha = 0$. In this case we immediately recover the standard Einstein relations for the separate charge and heat transport: $\sigma = \chi D_+$ and $\kappa = c_\rho D_-$. More generally the relation between the conductivities and diffusion constants does not separate in this simple way.

The generalized Einstein relations above can also be obtained by Kubo formulae. Following Kadanoff and Martin \cite{kadanoff,forster}, the diffusion equations (\ref{eq:dif}) imply that the spectral functions for $n_A$ at long wavelength satisfy
\be\label{eq:wk}
\frac{1}{\w} \chi_{A B}''(\w,k) = \text{Re} \left[ \frac{1}{- i \w + D k^2} \right]_{AC} \rchi_{CB} \,.
\ee
Here $\rchi_{AB}$ are the components of the susceptibilities appearing in (\ref{eq:change}). The Kubo formula then gives the matrix of conductivities
\be
\sigma_{AB} = \lim_{\w \to 0} \lim_{k \to 0} \frac{\w}{k^2} \, \chi_{A B}''(\w,k) = D_{AC} \rchi_{CB} \,.
\ee
These are the conductivities for the currents $\{j,j^E\}$. Using (\ref{eq:jQ}) to obtain the heat current $j^Q$, we immediately recover the Einstein relations (\ref{eq:e1aa}) and (\ref{eq:e2aa}).

\subsection*{Screening}

In the presence of long ranged Coulomb interactions, the longitudinal conductivity is given in terms of the charge density retarded Green's function $\chi_{\rho\rho}(\w,k)$ by (e.g. \cite{martin})
\be
\sigma^L(\omega,k) = \frac{- i \w \chi_{\rho\rho}(\w,k)}{k^2 - \chi_{\rho\rho}(\w,k)} \,.
\ee
We have used natural units for the electric charge. Using the diffusive form (considering decoupled charge diffusion for simplicity)
\be\label{eq:dform}
\chi_{\rho\rho}(\w,k) = \frac{k^2 D \chi}{i \w - D k^2} \,,
\ee
gives
\be\label{eq:sigmaL}
\sigma^L(\omega,k) = \frac{- i \omega D \chi}{i \omega - D (k^2 + \chi)} \,.
\ee
At long wavelengths, charged excitations are now seen to decay exponentially fast in time rather than diffuse. This is the effect of screening. However, (\ref{eq:sigmaL}) does not capture the d.c. conductivity correctly, as it vanishes as $\w \to 0$. This occurs because there are non-singular terms that have not been kept in (\ref{eq:dform}). For the d.c. conductivity, these terms can be accounted for by adding a constant to (\ref{eq:sigmaL}): $\sigma^L \to \sigma^L + \sigma_\text{d.c.}$. The constant is uniquely fixed by the Kramers-Kronig relation
\be
\sigma_\text{d.c.} = \frac{1}{i \pi} \int_{-\infty}^\infty \frac{\text{Im} \, \sigma^L(\omega,k)}{\w} d\w = D \chi \,.
\ee
So we see that the Einstein relation remains intact in the presence of screening, even while charge no longer diffuses on the longest wavelengths.

\section*{Acknowledgements}

I have benefitted greatly from discussions with Aharon Kapitulnik, Gabi Kotliar, Bob Laughlin, Andy Mackenzie, Ross McKenzie, Vadim Oganesyan, Joe Orenstein, Boris Spivak and especially Steve Kivelson. SAH is partially funded by a DOE Early Career Award and by a Sloan fellowship.


\begin{thebibliography}{99}

  \bibitem{andy}
 J.~A.~N.~Bruin, H.~Sakai, R.~S.~Perry and A.~P.~Mackenzie
``Similarity of Scattering Rates in Metals Showing T-Linear Resistivity,''
Science {\bf 339}, 804 (2013).

  \bibitem{subir}
 S.~Sachdev, {\it Quantum phase transitions}, CUP, 1999.

 \bibitem{jan}
 J.~Zaanen, ``Superconductivity:  Why the temperature is high,''
 Nature {\bf 430}, 512 (2004).

 \bibitem{hussey}
  N.~E.~Hussey, K.~Takenaka and H.~Takagi,
  ``Universality of the Mott-Ioffe-Regel limit in metals,''
  Phil. Mag. {\bf 84}, 2847 (2004) [cond-mat/0404263].

\bibitem{gun}
O.~Gunnarsson, M.~Calandra and J.~E.~Han. ``Colloquium: Saturation of electrical resistivity,''
Rev. Mod. Phys. {\bf 75}, 1085 (2003) [cond-mat/0305412].

\bibitem{Emery:1995zz} 
  V.~J.~Emery and S.~A.~Kivelson,
  ``Superconductivity in Bad Metals,''
  Phys.\ Rev.\ Lett.\  {\bf 74}, 3253 (1995).
  
\bibitem{Policastro:2001yc} 
  G.~Policastro, D.~T.~Son and A.~O.~Starinets,
  ``The Shear viscosity of strongly coupled N=4 supersymmetric Yang-Mills plasma,''
  Phys.\ Rev.\ Lett.\  {\bf 87}, 081601 (2001)
  [hep-th/0104066].
  
\bibitem{Herzog:2007ij} 
  C.~P.~Herzog, P.~Kovtun, S.~Sachdev and D.~T.~Son,
  ``Quantum critical transport, duality, and M-theory,''
  Phys.\ Rev.\ D {\bf 75}, 085020 (2007)
  [hep-th/0701036].
  
\bibitem{Hartnoll:2007ih} 
  S.~A.~Hartnoll, P.~K.~Kovtun, M.~Muller and S.~Sachdev,
  ``Theory of the Nernst effect near quantum phase transitions in condensed matter, and in dyonic black holes,''
  Phys.\ Rev.\ B {\bf 76}, 144502 (2007)
  [arXiv:0706.3215 [cond-mat.str-el]].
  
\bibitem{Kovtun:2004de} 
  P.~Kovtun, D.~T.~Son and A.~O.~Starinets,
  ``Viscosity in strongly interacting quantum field theories from black hole physics,''
  Phys.\ Rev.\ Lett.\  {\bf 94}, 111601 (2005)
  [hep-th/0405231].
  
\bibitem{Son:2007vk} 
  D.~T.~Son and A.~O.~Starinets,
  ``Viscosity, Black Holes, and Quantum Field Theory,''
  Ann.\ Rev.\ Nucl.\ Part.\ Sci.\  {\bf 57}, 95 (2007)
  [arXiv:0704.0240 [hep-th]].
  
\bibitem{Cremonini:2011iq} 
  S.~Cremonini,
  ``The Shear Viscosity to Entropy Ratio: A Status Report,''
  Mod.\ Phys.\ Lett.\ B {\bf 25}, 1867 (2011)
  [arXiv:1108.0677 [hep-th]].
  
\bibitem{Adams:2012th} 
  A.~Adams, L.~D.~Carr, T.~SchŠfer, P.~Steinberg and J.~E.~Thomas,
  ``Strongly Correlated Quantum Fluids: Ultracold Quantum Gases, Quantum Chromodynamic Plasmas, and Holographic Duality,''
  New J.\ Phys.\  {\bf 14}, 115009 (2012)
  [arXiv:1205.5180 [hep-th]].

\bibitem{ziman}
J.~M.~Ziman, {\it Electrons and phonons}, OUP, 1960.

\bibitem{forster}
D.~Forster, {\it Hydrodynamic Fluctuations, Broken Symmetry, and Correlation Functions},
W. A. Benjamin, Advanced Book Classics, 1975.

\bibitem{Hartnoll:2012rj} 
  S.~A.~Hartnoll and D.~M.~Hofman,
  ``Locally Critical Resistivities from Umklapp Scattering,''
  Phys.\ Rev.\ Lett.\  {\bf 108}, 241601 (2012)
  [arXiv:1201.3917 [hep-th]].

\bibitem{Mahajan:2013cja} 
  R.~Mahajan, M.~Barkeshli and S.~A.~Hartnoll,
  ``Non-Fermi liquids and the Wiedemann-Franz law,''
  Phys.\ Rev.\ B {\bf 88}, 125107 (2013)
  [arXiv:1304.4249 [cond-mat.str-el]].
  
  \bibitem{rosch}
  P.~Jung and A.~Rosch,
  ``Lower bounds for the conductivities of correlated quantum systems,''
  Phys.\ Rev.\ B {\bf 75}, 245104 (2007)
  [arXiv:0704.0886 [cond-mat.str-el]].

\bibitem{Hartnoll:2008hs} 
  S.~A.~Hartnoll and C.~P.~Herzog,
  ``Impure AdS/CFT correspondence,''
  Phys.\ Rev.\ D {\bf 77}, 106009 (2008)
  [arXiv:0801.1693 [hep-th]].

  \bibitem{elec}
   D. N. Basov, R. D. Averitt, D. van der Marel, M. Dressel and K. Haule,
  ``Electrodynamics of Correlated Electron Materials,''
   Rev. Mod. Phys. {\bf 83}, 471 (2011)
   [arXiv:1106.2309 [cond-mat.str-el]].
   
   \bibitem{fisher}
M.~P.~A.~Fisher, P.~B.~Weichman, G.~Grinstein, and D.~S.~Fisher,
``Boson localization and the superfluid-insulator transition,''
Phys. Rev. B {\bf 40}, 546 (1989).

  \bibitem{senthil1}
 T.~ Senthil,
``Theory of a continuous Mott transition in two dimensions,''
Phys. Rev. {\bf B78}, 045109 (2008) [arXiv:0804.1555 [cond-mat.str-el]].

\bibitem{senthil2}
W.~Witczak-Krempa, P.~Ghaemi, T.~Senthil, Y.~B.~Kim
``Universal transport near a quantum critical Mott transition in two dimensions,''
Phys. Rev. {\bf B86}, 245102 (2012) [arXiv:1206.3309 [cond-mat.str-el]].

\bibitem{Metlitski:2010vm}
  M.~A.~Metlitski and S.~Sachdev,
  ``Quantum phase transitions of metals in two spatial dimensions: II. Spin density wave order,''
  Phys.\ Rev.\ B {\bf 82}, 075128 (2010)
  [arXiv:1005.1288 [cond-mat.str-el]].
  
\bibitem{kad2}
R.~E.~Prange and L.~P.~Kadanoff,
``Transport theory for electron-phonon interactions in metals,"
Phys. Rev. {\bf 134}, A566 (1964).
  
  \bibitem{transverse}
 M.~Koschorreck, D.~Pertot, E.~Vogt and M.~Kšhl,
``Universal spin dynamics in two-dimensional Fermi gases,''
  Nat. Phys. {\bf 9}, 405 (2013).

\bibitem{Davison:2013txa} 
  R.~A.~Davison, K.~Schalm and J.~Zaanen,
  ``Holographic duality and the resistivity of strange metals,''
  arXiv:1311.2451 [hep-th].
  
\bibitem{Lucas:2014zea} 
  A.~Lucas, S.~Sachdev and K.~Schalm,
  ``Scale-invariant hyperscaling-violating holographic theories and the resistivity of strange metals with random-field disorder,''
  Phys.\ Rev.\ D {\bf 89}, 066018 (2014)
  [arXiv:1401.7993 [hep-th]].
  
    \bibitem{ks1}
  B.~Spivak and S.~A.~Kivelson,
  ``Transport in two dimensional electronic micro-emulsions,''
  Ann. Phys. {\bf 321}, 2071 (2006).

  \bibitem{ks2}
  A.~V.~Andreev, S.~A.~Kivelson and B.~Spivak
  ``Hydrodynamic description of transport in strongly correlated electron systems,''
  Phys. Rev. Lett. {\bf 106}, 256804 (2011).
  
\bibitem{Balasubramanian:2013yqa} 
  K.~Balasubramanian and C.~P.~Herzog,
  ``Losing Forward Momentum Holographically,''
  arXiv:1312.4953 [hep-th].
  
  \bibitem{dm1}
  A. Georges, G. Kotliar, W. Krauth and M. J. Rozenberg,
  ``Dynamical mean-field theory of strongly correlated fermion systems and the limit of infinite dimensions,''
  Rev. Mod. Phys. {\bf 68}, 13 (1996).
  
  \bibitem{dm2}
  X. Deng, J. Mravlje, R. \v{Z}itko, M. Ferrero, G. Kotliar and A. Georges,
  ``How Bad Metals Turn Good: Spectroscopic Signatures of Resilient Quasiparticles,''
  Phys. Rev. Lett. {\bf 110}, 086401 (2013).
  
  \bibitem{dmft}  
G.~P\'alsson and G.~Kotliar,
``Thermoelectric response near the density driven Mott transition,''
 Phys. Rev. Lett. {\bf 80}, 4775 (1998).
  
\bibitem{Kovtun:2008kx} 
  P.~Kovtun and A.~Ritz,
  ``Universal conductivity and central charges,''
  Phys.\ Rev.\ D {\bf 78}, 066009 (2008)
  [arXiv:0806.0110 [hep-th]].

\bibitem{ds}
K.~Damle and S.~Sachdev,
``Non-zero temperature transport near quantum critical points,''
 Phys. Rev. B {\bf 56}, 8714 (1997) [arXiv:cond-mat/9705206 [cond-mat.str-el]].

\bibitem{basov}  
  D. N. Basov and A. V. Chubukov,
  ``Manifesto for a higher Tc,''
 Nat. Phys. {\bf 7}, 272 (2011).
 
 \bibitem{MIT}
  M. Imada, A. Fujimori, Y. Tokura,
  ``Metal-insulator transitions,''
  Rev. Mod. Phys. {\bf 70}, 1039 (1998).
  
\bibitem{MIT2} 
G. Kotliar, S. Murthy and M. J. Rozenberg,
``Compressibility Divergence and the Finite Temperature Mott Transition,''
 Phys. Rev. Lett. {\bf 89}, 046401 (2002). 
 
\bibitem{MIT3}
J. Kokalj and R. H. McKenzie, 
``Thermodynamics of a Bad MetalÐMott Insulator Transition in the Presence of Frustration,''
Phys. Rev. Lett. {\bf 110}, 206402 (2013).
  
  \bibitem{ando}
 Y.~Ando, A.~N.~Lavrov, S.~Komiya, K.~Segawa and X.~F.~Sun,
 ``Mobility of the Doped Holes and the Antiferromagnetic Correlations in Underdoped High-Tc Cuprates,''
 Phys. Rev. Lett. {\bf 87}, 017001 (2001)
 [arXiv:cond-mat/0104163 [cond-mat.supr-con]].
  
  \bibitem{hussey1}
  N. E. Hussey, R. A. Cooper, Xiaofeng Xu, Y. Wang, B. Vignolle, C. Proust,
  ``Dichotomy in the T-linear resistivity in hole-doped cuprates,''
  Phil. Trans. R. Soc. A {\bf 369}, 1626 (2011)
  [arXiv:0912.2001 [cond-mat.supr-con]].
  
  \bibitem{orenstein}
  J. Orenstein, G. A. Thomas, A. J. Millis, S. L. Cooper, D. H. Rapkine, T. Timusk, L. F. Schneemeyer, and J. V. Waszczak, ``Frequency- and temperature-dependent conductivity in YBa$_2$Cu$_3$O$_{6+x}$ crystals,''
Phys. Rev. B {\bf 42}, 6342 (1990). 
  
  \bibitem{uchida}
  S. Uchida, T. Ido, H. Takagi, T. Arima, Y. Tokura, and S. Tajima,
  ``Optical spectra of La$_{2-x}$Sr$_x$CuO$_4$: Effect of carrier doping on the electronic structure of the CuO$_2$ plane,'' Phys. Rev. B {\bf 43}, 7942 (1991).
   
  \bibitem{taillefer}
  L. Taillefer,
  ``Scattering and Pairing in Cuprate Superconductors,''
  Ann. Rev. Cond. Mat. Phys. {\bf 1}, 51 (2010)
  [arXiv:1003.2972 [cond-mat.supr-con]].

\bibitem{takenaka}
K. Takenaka, J. Nohara, R. Shiozaki and S. Sugai,
``Incoherent charge dynamics of La$_{2-x}$Sr$_x$CuO$_4$: Dynamical localization and resistivity saturation,''
Phys. Rev. B {\bf 68}, 134501 (2003).

\bibitem{boris}
A. V. Boris, N. N. Kovaleva, O. V. Dolgov, T. Holden, C. T. Lin, B. Keimer and C. Bernhard,
``In-Plane Spectral Weight Shift of Charge Carriers in YBa$_2$Cu$_3$O$_{6.9}$,''
Science {\bf 304}, 708 (2004).

\bibitem{marel}
D. van der Marel, H. J. A. Molegraaf, J. Zaanen, Z. Nussinov, F. Carbone, A. Damascelli, H. Eisaki, M. Greven, P. H. Kes and M. Li
``Quantum critical behaviour in a high-T$_c$ superconductor,''
Nature {\bf 425}, 271 (2003).

\bibitem{hwang}
J Hwang, T Timusk and G D Gu,
``Doping dependent optical properties of Bi$_2$Sr$_2$CaCu$_2$O$_{8+\delta}$,''
J. Phys.: Cond. Mat. {\bf 19}, 125208 (2007).

\bibitem{p1}
D. Wu, N. Barisic, N. Drichko, S. Kaiser, A. Faridian, M. Dressel, S. Jiang, Z. Ren, L. J. Li, G. H. Cao, Z. A. Xu, H. S. Jeevan and P. Gegenwart,
``Effects of magnetic ordering on dynamical conductivity: Optical investigations of EuFe$_2$As$_2$ single crystals,''
Phys. Rev. B {\bf 79}, 155103 (2009).

\bibitem{p2}
 A. A. Schafgans, S. J. Moon, B. C. Pursley, A. D. LaForge, M. M. Qazilbash, A. S. Sefat, D. Mandrus, K. Haule, G. Kotliar and D. N. Basov,
 ``Electronic Correlations and Unconventional Spectral Weight Transfer in the High-Temperature Pnictide BaFe$_{2-x}$Co$_x$As$_2$ Superconductor Using Infrared Spectroscopy,''
 Phys. Rev. Lett. {\bf 108}, 147002 (2012).
 
 \bibitem{i1}
  K. Takenaka, M. Tamura, N. Tajima, H. Takagi, J. Nohara and S. Sugai,
  ``Collapse of Coherent Quasiparticle States in $\theta$-(BEDT-TTF)$_2$I$_3$ Observed by Optical Spectroscopy,''
    Phys. Rev. Lett. {\bf 95}, 227801 (2005). 
 
 \bibitem{i2} 
 P. E. J\"onsson, K. Takenaka, S. Niitaka, T. Sasagawa, S. Sugai and H. Takagi, 
``Correlation-Driven Heavy-Fermion Formation in LiV$_2$O$_4$,''
Phys. Rev. Lett. {\bf 99},167402 (2007).

\bibitem{i3}
 N. L. Wang, P. Zheng, D. Wu, Y. C. Ma, T. Xiang, R. Y. Jin and D. Mandrus,
 ``Infrared Probe of the Electronic Structure and Charge Dynamics of Na$_{0.7}$CoO$_2$,''
 Phys. Rev. Lett. {\bf 93}, 237007 (2004).
 
 \bibitem{i4}
  Y. S. Lee, Jaejun Yu, J. S. Lee, T. W. Noh, T.-H. Gimm, Han-Yong Choi and C. B. Eom,
  ``Non-Fermi liquid behavior and scaling of the low-frequency suppression in the optical conductivity spectra of CaRuO$_3$,''  Phys. Rev. B {\bf 66}, 041104(R) (2002).
  
\bibitem{WitczakKrempa:2012gn} 
  W.~Witczak-Krempa and S.~Sachdev,
  ``The quasi-normal modes of quantum criticality,''
  Phys.\ Rev.\ B {\bf 86}, 235115 (2012)
  [arXiv:1210.4166 [cond-mat.str-el]].
 
\bibitem{kot} 
X. Deng, A. Sternbach, K. Haule, D. N. Basov and G. Kotliar,
``Shining light on transition metal oxides: unveiling the hidden Fermi Liquid,''
 arXiv:1404.6480 [cond-mat.str-el].
 
\bibitem{nickel} 
R. Jaramillo, S. D. Ha, D. M. Silevitch and	S. Ramanathan,
``Origins of bad-metal conductivity and the insulatorÐmetal transition in the rare-earth nickelates,''
Nat. Phys. {\bf 10}, 304 (2014).
 
 \bibitem{tanatar}
M. A. Tanatar, J. Paglione, C. Petrovic, L. Taillefer 
``Anisotropic Violation of the Wiedemann-Franz Law at a Quantum Critical Point,''
Science {\bf 316}, 1320 (2007).

\bibitem{con1}
H.~Pfau, S.~Hartmann, U.~Stockert,	P.~Sun, S.~Lausberg, M.~ Brando, S.~Friedemann,	C.~Krellner, C.~Geibel, S.~Wirth, S.~Kirchner, E.~Abrahams, Q.~Si and F.~Steglich,
``Thermal and electrical transport across a magnetic quantum critical point,''
Nature {\bf 484}, 493 (2012).

\bibitem{zhang}
Y.~Zhang, N.~P.~Ong, Z.~A.~Xu, K.~Krishana, R.~Gagnon and L.~Taillefer,
``Determining the Wiedemann-Franz Ratio from the Thermal Hall Conductivity: Application to Cu and YBa$_{2}$Cu$_{3}$O$_{6.95}$,''
Phys. Rev. Let. {\bf 84}, 2219 (2000).

\bibitem{d1}
    X.D. Wu, J.G. Fanton, G.S. Kino, S. Ryu, D.B. Mitzi and A. Kapitulnik
 ``Thermal diffusivity of Bi$_2$Sr$_2$CaCu$_2$O$_8$ single crystals,''
 Physica C: Super. {\bf 218}, 417 (1993).
 
 \bibitem{d2} 
 X. D. Wu, G. S. Kino, J. T. Fanton and A. Kapitulnik,
 ``Photothermal microscope for high-Tc superconductors and charge density waves,''
 Rev. Sci. Instrum. {\bf 64}, 3321 (1993).
 
 \bibitem{spin}
 C. P. Weber, N. Gedik, J. E. Moore, J. Orenstein, J. Stephens  and D. D. Awschalom,
 ``Observation of spin Coulomb drag in a two-dimensional electron gas,''
 Nature {\bf 437}, 1330 (2005).
 
 \bibitem{qp}
 N. Gedik, J. Orenstein, Ruixing Liang, D. A. Bonn and W. N. Hardy,
 ``Diffusion of Nonequilibrium Quasi-Particles in a Cuprate Superconductor,''
Science {\bf 300}, 1410 (5624).
 
\bibitem{Donos:2014uba} 
  A.~Donos and J.~P.~Gauntlett,
  ``Novel metals and insulators from holography,''
  arXiv:1401.5077 [hep-th].

\bibitem{Gouteraux:2014hca} 
  B.~Gout\'eraux,
  ``Charge transport in holography with momentum dissipation,''
  arXiv:1401.5436 [hep-th].
  
\bibitem{Karch:2007pd} 
  A.~Karch and A.~O'Bannon,
  ``Metallic AdS/CFT,''
  JHEP {\bf 0709}, 024 (2007)
  [arXiv:0705.3870 [hep-th]].
  
\bibitem{Hartnoll:2009ns} 
  S.~A.~Hartnoll, J.~Polchinski, E.~Silverstein and D.~Tong,
  ``Towards strange metallic holography,''
  JHEP {\bf 1004}, 120 (2010)
  [arXiv:0912.1061 [hep-th]].
  
    \bibitem{vadim} 
S. Mukerjee, V. Oganesyan and D. Huse,
``Towards a statistical theory of transport by strongly-interacting lattice fermions,''
 Phys. Rev. B {\bf 73}, 035113 (2006)
 [arXiv:cond-mat/0503177 [cond-mat.str-el]].
  
  \bibitem{kadanoff}
L.~P.~Kadanoff and P.~C.~Martin,
``Hydrodynamic equations and correlation functions,''
Ann. Phys., {\bf 24}, 419 (1963).

\bibitem{martin}
P.~C.~Martin, {\it Measurements and correlation functions}, Gordon and Breach, 1968.

\end{thebibliography}
\end{document}